\documentclass[12pt]{article}
\usepackage{graphicx}\usepackage{amssymb}\usepackage{bm}\usepackage{amsfonts}\usepackage{amsmath}\usepackage{cases}
\usepackage{wrapfig} 

\newcommand{\beq}{\begin{equation}}
\newcommand{\eeq}{\end{equation}}
\newcommand{\beqa}{\begin{eqnarray}}
\newcommand{\eeqa}{\end{eqnarray}}\newcommand{\w}{\wedge}
\newcommand{\dl}{\bm{\delta}}\newcommand{\nn}{\nonumber}

\newcommand{\h}[1]{\hat{#1}}

\newcommand{\ou}[3]{\underset{#3}{\overset{#1}{#2}}}

\newcommand{\ua}{\uparrow}
\newcommand{\da}{\downarrow}

\newcommand{\g}{\gamma}

\newcommand{\p}{\partial}

\newcommand{\mc}{\mathcal}
\newcommand{\bb}{\mathbb}
\newcommand{\mf}{\mathfrak}
\newcommand{\cA}{\mathcal{A}}
\newcommand{\hI}{\hat{I}}
\newcommand{\hJ}{\hat{J}}
\newcommand{\hK}{\hat{K}}
\newcommand{\hL}{\hat{L}}
\newcommand{\hM}{\hat{M}}

\begin{document}
{\renewcommand{\thefootnote}{\fnsymbol{footnote}}

\begin{center}
{\LARGE Is de Sitter space a fermion?}\\
\vspace{1.5em}
Andrew Randono\footnote{e-mail address: {\tt arandono@gmail.com}}
\\
\vspace{0.5em}
The Perimeter Institute for Theoretical Physics \\
31 Caroline Street North\\
Waterloo, ON N2L 2Y5, Canada
\vspace{1.5em}

\vspace{0.5em}
W.M. Keck Science Center \\
Claremont College Consortium\\
925 N. Mills Ave\\
Claremont, CA 91711-5916
\vspace{1.5em}
\end{center}
}

\setcounter{footnote}{0}

\begin{abstract}
Following up on a recent model yielding fermionic geometries, I turn to more familiar territory to address the question of statistics in purely geometric theories. Working in the gauge formulation of gravity, where geometry is characterized by a symmetry broken Cartan connection, I give strong evidence to suggest that de Sitter space itself, and a class of de Sitter-like geometries, can be consistently quantized fermionically. By this I mean that de Sitter space can be quantized such that the wavefunctional picks up an overall minus sign under a $2\pi$ rotational diffeomorphism. Surprisingly, the underlying mathematics is the same as that of the Skyrme model for strongly interacting baryons. This promotes the question {\it ``Is geometry bosonic or fermionic?"} beyond the realm of the rhetorical and places it on uncomfortably familiar ground.
\end{abstract}

\section{Introduction} 

It is generally taken on faith that the geometry underlying general relativity is bosonic. In quantum gravity, this  faith is buttressed by the observation that low energy, small-amplitude excitations of the gravitational field are spin-2 gravitons, and should therefore be quantized as bosons. On the other hand, it is well known that certain constrained or otherwise non-linear field theories whose low energy excitations are bosonic, can nevertheless give rise to emergent structures in the non-perturbative regime with fermionic (or anyonic) statistics \cite{Arnsdorf:1998vq,Finkelstein:1968hy,Giulini:1993gd,Skyrme:1961vq,Skyrme:1962vh,Skyrme:Original,Skyrme:Original2,Solitons,TopologicalSolitons,Williams:Skyrme}. It is conceivable then that the phase space structure of general relativity could admit fermionic modes upon quantization.

To support this hypothesis, in a recent article we constructed a purely geometric theory with precisely this peculiar property \cite{Randono:2010cd,Randono:2011bb}. The model presented there is not general relativity, however it is a dynamical geometry theory.
Specifically, we considered a propagating torsion theory in flat Minkowski space, where the torsion is constrained in such a way that the local degrees of freedom of the model are described by a non-linear sigma model with target space $Spin(3,1)$. By formally mapping the model onto the Skyrme model for strongly interacting baryons, we showed that the system could  be quantized in such a way that isolated torsional charges of even charge behave as bosons under rotations and exchanges, whereas odd charges behave as fermions. 

The existence of these fermionic geometries leads one to question whether this could be a generic feature of dynamical geometries or simply a peculiarity of the particular model. One may raise the speculative objection, for example, that the possibility of more exotic non-pertubative statistics is novel to torsional theories and cannot necessarily be extrapolated to non-torsional geometries. In this paper I will give strong evidence to quell this objection.

Specifically I will show that a much more familiar geometry, namely de Sitter space itself and a class of generalized de Sitter-like geometries, can be quantized fermionically. The geometric arena I will work in is the gauge formulation of gravity where geometry is described by a reductive Cartan connection on a $Spin(4,1)$-bundle \cite{Randono:Condensate,Randono:dSSpaces,Utiyama:1973nq,Utiyama:1980bp,West:1978Lagrangian,Wise:2009fu,Wise:MMGravity,Randono:Review}. Surprisingly, the underlying mathematics allowing for fermionic quantization is the same as for the Skyrme model. 

Bosonic geometry has long been an implicit tenet of quantum gravity theories. The possibility of fermionic geometries will likely require significant rethinking of foundational issues in quantum geometry. Furthermore, de Sitter space plays an important role in cosmology as the expected ground state of a universe with a positive cosmological constant and potentially as the future asymptote of our universe. This promotes the question {\it ``Is geometry bosonic or fermionic?"} beyond the realm of the rhetorical and places it on uncomfortably familiar ground.

\subsection{The results, briefly}
Using the term {\it fermionic} to describe an object that is not pointlike (or even more amusingly, as is the case here, a full geometry) is likely to be unfamiliar to many readers. However, it has been known for some time that extended objects can have topological properties that consistently reproduce the familiar properties of fermionic spinors under rotations and exchanges. The necessary topological property that such an object must possess is that the transformation given by a parameterized $2\pi$-rotation cannot be homotopic to the identity map. Rather, it serves as the homotopy generator of a $\bb{Z}_2$ map (since $4\pi$ rotations {\it are} homotopic to the identity). This allows one to extend the concept of fermionic statistics to full geometries. As I will show, the geometry describing de Sitter space can be described in almost the same mathematical framework as Skyrmions, which opens the door to fermionic quantization in the same way that Skyrmions do.

The transformation that will serve as the generator of the $\bb{Z}_2$ degeneracy is itself a diffeomorphism serving as a time-parameterized rotation of the spacetime by $2\pi$. At this point, certain obvious objections to the relevance of this framework to quantum gravity need to be addressed. Specifically, since the transformation in question is a diffeomorphism, and diffeomorphisms are generally believed to act trivially on the physical quantum Hilbert space, it does not seem possible that the fermionic nature of the geometry, characterized by $\Psi$ transforming to $-\Psi$ under a $2\pi$ rotation, could be captured by an ordinary diffeomorphism. 

There are two retorts to this objection. First, the triviality of diffeomorphisms on the physical Hilbert space of quantum gravity follows from the vanishing of the diffeomorphism constraint, which itself is the generator of {\it identity connected} diffeomorphisms. However, as I will show, the diffeomorphism representing a parameterized $2\pi$-rotation is not identity connected. Second, the action of quantum operators that {\it usually} annihilate the physical wavefunctional can be more subtle when the phase space of the theory has non-trivial cohomology. In the geometric quantization framework,  non-triviality of the first cohomology group allows for a quantization ambiguity stemming from topologically distinct symplectic one-forms giving rise to the same symplectic structure. In fact, I will construct a quantum theory from a phase space that is at once rich enough to contain de Sitter space in a full quantum Hilbert space, yet simple enough highlight the interesting and relevant topological structures. As I will show, the phase space of this theory has precisely this property. The non-trivial cohomology allows for two distinct quantizations: one in which diffeomorphisms act trivially on the physical Hilbert space, and one in which diffeomorphisms act {\it projectively} on the physical Hilbert space. In the latter case, the action of a diffeomorphism is to send $\Psi$ to $e^{i\theta}\Psi$ (this is what is meant by {\it projective}). This representation we will refer to as {\it fermionic quantization} of de Sitter space, because in this representation the phase conspires to send $\Psi$ to $-\Psi$ under a time-parameterized $2\pi$-rotational diffeomorphism.

\section{Geometry from Cartan's perspective}
The mathematical arena that I will work in is the reformulation of Einstein-Cartan theory as a symmetry broken gauge theory. I will briefly review this formalism below, but I refer the reader to my review paper \cite{Randono:Review} for a more extensive presentation. 

In the gauge formulation of gravity, geometry is characterized by a Cartan connection $\cA$ taking values in a reductive Cartan algebra $\mf{g}=\mf{h}\oplus\mf{p}$. The reductive split of the Lie algebra into a stabilizer subalgebra $\mf{h}$ and its complement $\mf{p}$ is facilitated through the introduction \cite{Stelle:1979aj,Stelle:1979va} of a new field $V$. This field acts as a symmetry breaking field analogous to the Higgs, ``breaking'' the $\mc{G}$ symmetry of the principle $\mc{G}$-bundle down to a subgroup $\mc{H}$ which stabilizes $V$ at each point. The field itself takes values in a space isomorphic to the coset space $\mc{G}/\mc{H}$. Thus, in total the geometry is characterized by a specification of the pair $\{\cA,V\}$, modulo $\mc{G}$-gauge transformations. The breaking of the symmetry allows for a separation of the spin connection from the frame field corresponding to the reductive split $\mf{g}=\mf{h}\oplus \mf{p}$, so that $\cA=\omega \oplus \frac{1}{\ell}e$ where $\ell$ is a parameter with dimension of length. Here, $\omega$ is the $\mf{h}$-valued spin-connection, and $e$ is the $\mf{p}$-valued frame field. For the case of gravity in $(3+1)$-dimensions with a positive cosmological constant, the gauge group is $\mc{G}=Spin(4,1)$, the double cover of the de Sitter group, and the stabilizer subgroup is $\mc{H}=Spin(3,1)$. This implies that the symmetry breaking field $V$ takes values in the coset space $\mc{G}/\mc{H}\simeq \bb{R}\times\bb{S}^3$. The parameter $\ell$ can be related to the cosmological constant $\Lambda$ by $\ell=\sqrt{3/\Lambda}$.

To see this more explicitly, consider the adjoint representation of $\mc{G}=Spin(4,1)$. Let hatted-upper case Roman indices $\{\h{I}, \h{J}, \h{K},\dots\}$ range from $0$ to $4$, and unhatted indices $\{I,J,K,\dots\}$ range from $0$ to $3$. The symmetry breaking field $V^{\h{I}}$ is a vector representation of $Spin(4,1)$ whose magnitude is constrained by $\eta_{\h{I}\h{J}}V^{\h{I}}V^{\h{J}}=1$ with $\eta_{\h{I}\h{J}}=diag(-1,1,1,1,1)$. The lift of the tetrad to the $Spin(4,1)$ vector space can then be identified with $e^{\h{I}}\equiv \ell D_{\mc{A}} V^{\h{I}}$ and the spin connection with $\omega^{\h{I}\h{J}}=\cA^{\h{I}\h{J}}-2D_\cA V^{[\h{I}} \,V^{\h{J}]}$. 

One can always choose a specific gauge locally where $V^{\h{I}}=(0,0,0,0,1)$. For the rest of the paper we will refer to this gauge as the {\bf Einstein-Cartan (EC) gauge}. In the EC gauge, the tetrad simplifies to $e^I=\ell \cA^{I4}$, and the spin connection to $\omega^{IJ}=\cA^{IJ}$. In a Clifford algebra notation which we will employ here (see Appendix A) the symmetry breaking field is given by $V=V_{\h{I}}\g^{\h{I}}$ with $\g^{\h{I}}=(\g^I,\g_5)$. The action of a $Spin(4,1)$ gauge transformation generated by $g=g(x)$ is given by $\{\cA,V\}\rightarrow \{g\cA g^{-1}-dg\,g^{-1},gVg^{-1}\}$.

Corresponding to the split\footnote{It is convenient in the Clifford algebra notation to define $e=\frac{1}{2}\g_I\, e^I$ which pulls the $\g_5$ out of the expression in the decomposition. See Appendix A for more details.} $\cA=\omega+ \frac{1}{\ell}\g_5 e$, the curvature also splits into 
\beq
\mc{F}_\cA=\underbrace{R_{\omega}-\frac{1}{\ell^2} e\w e}_{\mf{h}}\ \  + \ \ \underbrace{\g_5 \frac{1}{\ell}T}_{\mf{p}}\,.
\eeq
The constant curvature ($R_{\omega}=\frac{1}{\ell^2} e\w e$), zero torsion ($T=0$) condition is then given succinctly by $\mc{F}_{\mc{A}}=0$. Thus, in the gauge framework of gravity, de Sitter space is described by a flat connection. This allows one to easily embed the de Sitter solution into a larger space, which is at the same time small enough to easily characterize topologically. For this reason, I will focus on the configurations consisting of the pair $\{\cA, V\}$ where $\cA$ is restricted to be a flat connection. I will refine this phase space shortly.

\subsection{Flat Cartan geometries}
Restrict attention to manifolds with the topology of de Sitter space, namely $M\simeq \bb{R}\times \bb{S}^3$. Since $\pi_1(M)=0$, all flat connection are ``gauge related" to the zero connection. By this I mean that there exists a $g:\bb{R}\times \bb{S}^3 \rightarrow Spin(4,1)$ such that $\cA=-dg g^{-1}$. The geometry is therefore characterized by the pair $\{\cA,V\}=\{-dg\,g^{-1},V\}$. However, one can always perform a {\it true} gauge transformation (which transforms {\it both} fields simultaneously) such that after the transformation we have $\cA=0$ and $V'=g^{-1}Vg$, thereby pushing all the geometric information into the symmetry breaking field itself. In this gauge, which I will refer to as the {\bf trivial gauge}, the geometry is completely characterized by the map $V:\bb{R}\times \bb{S}^3\rightarrow \mc{G}/\mc{H} \simeq \bb{R}\times \bb{S}^3$. Because of this, the trivial gauge will be extremely useful in analyzing the topological properties of the geometries in this framework.

The goal is to categorize the space of flat Cartan connections topologically. To this end, consider first the set of maps that are deformable to a time independent map $V:\bb{S}^3 \rightarrow \mc{G}/\mc{H}\simeq \bb{R}\times \bb{S}^3$. The homotopy class of maps of this sort fall into discrete classes characterized by $\pi_3(\mc{G}/\mc{H})=\bb{Z}$. The integer labelling the sector in which the map resides is the number of times the $\bb{S}^3$ of the range winds around the $\bb{S}^3$ of the domain, or the winding number of the map.

As shown in \cite{Randono:dSSpaces, Randono:Review} the $\bb{Z}$-sectors of the phase space can be constructed as follows. Define $\ou{m}{h}{n}$ to be the time-independent map $\ou{m}{h}{n}:\bb{S}^3 \rightarrow Spin(4,1)$ given by
\beq
\ou{m}{h}{n}=\left[ \begin{matrix} h^m & 0 \\ 0 & h^n    \end{matrix} \right]  \quad \mbox{with} \quad h=X^4 \bm{1}+X^i \,i\sigma_i  \label{Zgenerator}
\eeq
where $X$ denotes the usual embedding of the unit three sphere in $\bb{R}^4$ given explicitly in three-dimensional polar coordinates by
\beqa
X^1 &=& \sin\chi \,\sin{\theta} \,\cos\phi \nn\\
X^2 &=& \sin\chi \, \sin{\theta}\,\sin{\phi} \nn\\
X^3 &=& \sin\chi \, \cos{\theta} \nn\\
X^4 &=& \cos{\chi}
\eeqa

 The field $h$ is a map $h:\bb{S}^3 \rightarrow SU(2)$ with winding number one, and as such it is the generator of $\pi_3(SU(2))=\bb{Z}$. Since $Spin(4) \simeq SU(2)_\ua \times SU(2)_\da$ and $Spin(4)$ is the maximal compact subgroup of $Spin(4,1)$, $\ou{m}{h}{n}(x)\in Spin(4,1)$ and it has winding number $p\equiv m+n$. To add back in the time dependence of the de Sitter solutions, first time-translate $\ou{m}{h}{n}$ to $\ou{m}{g}{n}=h_t \ou{m}{h}{n}$ where $h_t=\exp(\frac{1}{2}t \g_5 \g^0)$. A class of topologically distinct Cartan geometries is then given, in the EC gauge, by the configurations
\beq
\{\cA,V\}=\{\ou{m}{\cA}{n} \equiv-d\ou{m}{g}{n}\,\ou{m}{g}{n}{}^{-1}, \g_5\}\,. \label{AConfig}
\eeq

To understand the topological structure of these configurations more thoroughly it is useful to transform to the trivial gauge where $\cA=0$. In this gauge, it can be shown that $V'=\ou{m}{V}{n}\equiv \ou{m}{g}{n}{}^{-1} \g_5 \ou{m}{g}{n}{}^{-1}$ viewed as a map $\ou{m}{V}{n}:\bb{R}\times \bb{S}^3 \rightarrow \mc{G}/\mc{H}$ has winding number $-q$ where $q\equiv m-n$. Thus, the integer $q$ delineates the topological sectors of the space of flat Cartan connections corresponding to the homotopy group $\pi_3(\mc{G}/\mc{H})$.

\begin{wrapfigure}{r}{0.5\textwidth}
  \begin{center}
    \includegraphics[width=0.4\textwidth]{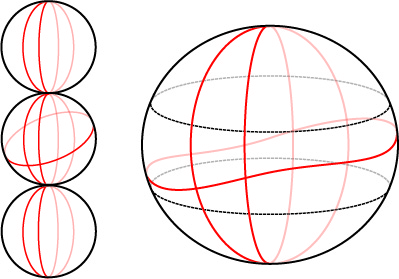}
  \end{center}
  \caption{\label{3Spheres} A visualization of the generalized de Sitter space corresponding to $|q|=3$ on a constant time slice. On the left is the standard visualization where the radius is weighted by volume. The geometry consists of a string of $|q|$ three-spheres attached at the poles along two-spheres where the metric becomes degenerate. On the right is a more topologically accurate picture where the degenerate surfaces are pictured as extended regions denoted by dotted lines.}
\end{wrapfigure}

Geometrically these configurations have a very simple interpretation. The metric induced from the pair $\{\ou{m}{\cA}{n},V=\g_5\}$ is given by
\beq
\ou{m}{\bm{g}}{n}=-dt^2 +\cosh^2(t/\ell)\left(|q|^2 d\chi^2 +\sin^2(|q| \chi) \left(d\theta^2+\sin^2\theta \,d\phi^2 \right) \right)\,. \label{GConfig}
\eeq
For $|q| >0$, this metric describes a ``string of pearls" geometry consisting of $|q|$ copies of de Sitter space attached at their poles by regions where the metric becomes degenerate (see Fig. \ref{3Spheres}). The geometry of each individual 3-sphere is identical to that of de Sitter space itself aside from the poles. One can show that the oriented volume of the three-sphere defined at the throat of de Sitter space at $t=0$ is given by $2\pi^2 \ell^3 (-q)$.

It should be stressed that all these configurations are solutions to the Einstein Cartan field equations, despite being degenerate \cite{Randono:dSSpaces, Randono:Review}. This serves to generalize de Sitter space, and allows one to embed the pure de Sitter solution into a larger phase space that is still manageable to work with.  

\subsection{$\bb{Z}_2$ degeneracy of the phase space}
I will now show that the phase space containing the de Sitter configuration admits one further degeneracy. But first it is necessary to refine and define the phase space more precisely. Consider the set of flat Cartan geometries that asymptotically approach the pair $\{\ou{m}{\cA}{n},V=\g_5\}$ in the past and future. In the trivial gauge this can be simplified to the set of maps $V:\bb{R}\times \bb{S}^3 \rightarrow \mc{G}/\mc{H}$ that asymptote to $\ou{m}{V}{n}$. In this fixed gauge, the only remaining gauge invariance is a global action of a constant $g\in Spin(4,1)$. Rather than modding out by diffeomorphism equivalence classes, we will consider the action of diffeomorphisms on the phase space that preserves the geometry on the boundaries. This allows for a classification of the action of diffeomorphisms into types categorized by topological properties. Thus, I will take the phase space $\mc{Q}$ to be the space of flat Cartan connections modulo the set of {\it identity connected}, boundary isometries. 

The two lowest non-zero homotopy groups of the target space of $V$ are $\pi_3(\mc{G}/\mc{H})=\pi_3(\bb{S}^3)=\bb{Z}$ and $\pi_4(\mc{G}/\mc{H})=\pi_4(\bb{S}^3)=\bb{Z}_2$. I have already shown that the former is the topological property that allowed for the construction of a class of configurations labelled by the integer $q$. The remaining $\bb{Z}_2$ degeneracy I will argue allows for fermionic quantization of de Sitter space. In total, the space of flat Cartan connections $\mc{Q}$ splits into topologically distinct sectors labelled by the winding number, and the $\bb{Z}_2$ degeneracy.

\section{The action of diffeomorphisms on $\mc{Q}$}
The goal now is to construct a generator of the $\bb{Z}_2$ degeneracy. In fact, the degeneracy can be related to the action of diffeomorphisms on $\mc{Q}$. First, however, I will discuss generically how this degeneracy comes about. 

\begin{figure}
 \begin{center}
\includegraphics[height=6.0cm]{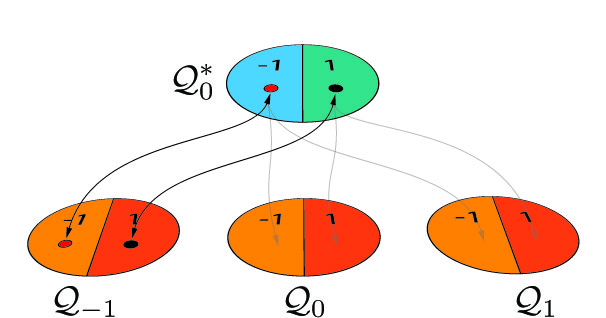}
   \end{center}
  \caption{\label{Exchange} \label{Homeomorphism} Schematic of the homeomorphism between the $\mc{Q}_q$ sectors and the reference space $\mc{Q}^*_0$. On the left is the $\mc{Q}_{-1}$ sector containing the de Sitter solution (black point) and the twisted de Sitter solution (red point) which are mapped to different $\bb{Z}_2$ sectors of $\mc{Q}^*_0$.  }
 \end{figure}

Start with the $q=0$ sector denoted $\mc{Q}_0$. In this sector, the configuration should asymptote to $\{\ou{0}{\cA}{0}=- dh_t\,h_t^{-1},\g_5\}$. As in \cite{Giulini:1993gd}, to analyze the topological properties, it is convenient to define a canonical homeomorphism between each topological sector and a space where the analytic properties of of the map are more apparent. Thus, for each sector $\mc{Q}_q$, we define a homeomorphism to a new sector $\mc{Q}^*_0$, which is itself homotopic to $\mc{Q}_0$. The homeomorphisms is defined as follows (see Fig. \ref{Homeomorphism}). Given a configuration in the $\mc{Q}_q$ sector first transform to the EC gauge where $V=\g_5$. Next map the pair $\{\cA, \g_5\}\rightarrow \{h_t^{-1}\cA h_t -dh_t^{-1} h_t, \g_5\}$. This serves to remove the time dependence of the fiducial configurations $\{\ou{m}{\cA}{n}, \g_5\}$.
Now transform to the trivial gauge. Finally transform the configuration by $\{0,V\}\rightarrow \{0,\ou{0}{h}{q} V \ou{0}{h}{q}{}^{-1}\}$. 

The point behind this homeomorphism is that it transforms each of the fiducial configurations $\{\ou{m}{\cA}{n},\g_5\}$ defined in the previous section to the convenient base point $\{0,\g_5\}$ in $\mc{Q}^*_0$. This will make it easier to determine the homotopy properties of the configuration.

The resulting configuration lives in a space $\mc{Q}_0^*$. This space is formed by the set of maps $V:\bb{R}\times \bb{S}^3 \rightarrow \mc{G}/\mc{H}$ that asymptote to $V=\g_5$ in the asymptotic past and future. Because of the latter restriction, one can add the endpoints $t=\{+\infty\}$ and $t=\{-\infty\}$ and compactify the domain to $\bb{S}^4$. Thus $V:\bb{S}^4 \rightarrow \mc{G}/\mc{H}$. The homotopy classes of maps of this type are characterized by $\pi_4(\mc{G}/\mc{H})=\bb{Z}_2$. Thus, via this homeomorphism, all states in $\mc{Q}$ can be classified into two groups: those configurations that when mapped to $\mc{Q}^*_0$ are homotopic to $\{0,\g_5\}$ and those that are homotopic to a generator of $\pi_4(\mc{G}/\mc{H})=\bb{Z}_2$ in $\mc{Q}^*_0$.


\subsection{Rotational diffeomorphisms and the $\bb{Z}_2$ degeneracy}
Now let us consider the action of diffeomorphisms on $\mc{Q}$. To preserve the phase space, one should restrict attention to those that asymptote to an isomorphism in the asymptotic past and future. However, since an isomorphism is equivalent to a gauge transformation by a constant element of $Spin(4,1)$ and is therefore contained in the local gauge group, we will restrict to $\mbox{\it Diff}_0$, the set of diffeomorphisms that tend to the identity at future and past asymptotic infinity.

\begin{wrapfigure}{r}{0.4\textwidth}
  \begin{center}
    \includegraphics[width=0.38\textwidth]{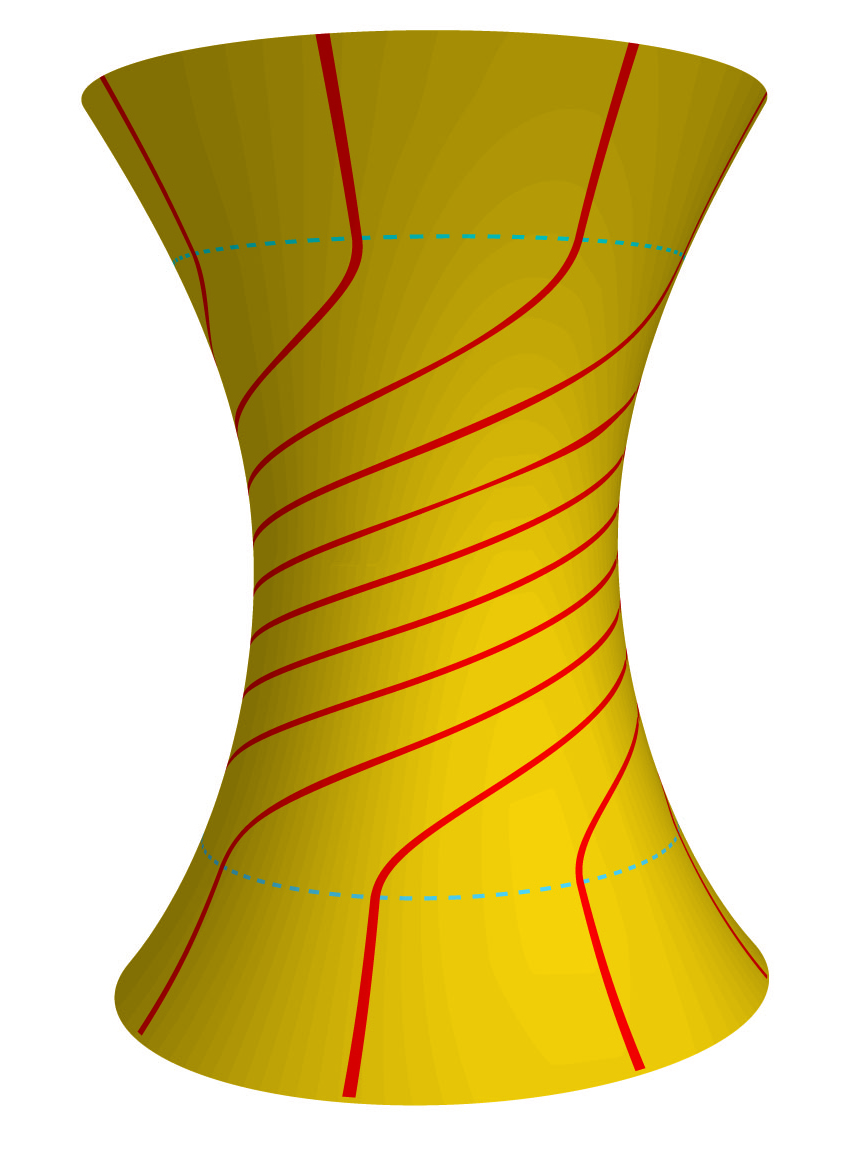}
  \end{center}
  \caption{\label{TwisteddS} The twisted version of de Sitter space given by the metric (\ref{TwisteddSMetric}). The red lines represent lines of constant $\phi'$.}
\end{wrapfigure}

Consider first the $q=-1$ sector where the fiducial configuration $\{\ou{0}{\cA}{1},\g_5\}$ represents ordinary de Sitter space. Define a one-parameter rotational diffeomorphism $\varphi$ by its action on the coordinates $\varphi(\{t,\chi, \theta, \phi\})=\{t,\chi,\theta,\phi'\equiv\phi-\phi_0(t)\}$ where $\phi_0(t)$ is a smooth function of $t$ only that varies from $0$ to $2\pi$ in the interval $[t_i,t_f]$, and is constant outside the interval. Assume also that the derivative $\p_t\phi_0$ vanishes at $t_i$ and $t_f$. For definiteness, choose $t_i=-\ell$ and $t_f=\ell$. Explicitly, the diffeomorphism is given by $\varphi=\exp(-\phi_0\mc{L}_{\bar{\phi}})$ where $\bar{\phi}=\frac{\p}{\p\phi}$, and its action on de Sitter space is represented by a $2\pi$ twist. The metric transforms to (see Fig. \ref{TwisteddS} for visualization)
\beq
\ou{0}{\bm{g}}{1}=-(1-\alpha)dt^2 +2\beta \,dt\,d\phi' +\cosh^2(t/\ell)\left(d\chi^2 +\sin^2\chi\left(d\theta^2+\sin^2\theta\, d\phi'{}^2\right)\right) \label{TwisteddSMetric}
\eeq
where $\alpha=\cosh^2(t/\ell)\,\sin^2\chi \,\sin^2\theta \,(\p_t\phi'_0)^2$ and $\beta=\cosh^2(t/\ell)\,\sin^2\chi \,\sin^2\theta \,\p_t\phi'_0$.


Consider now the action of of $\varphi$ on the configuration $\{\ou{0}{\cA}{1}, \g_5\}$. The diffeomorphism leaves $V=\g_5$ fixed in this gauge, but transforms $\ou{0}{\cA}{1}=-d\ou{0}{g}{1}\,\ou{0}{g}{1}{}^{-1}$ to $\varphi(\ou{0}{\cA}{1})=-d\ou{0}{g}{1}{}'\,\ou{0}{g}{1}{}'^{-1}$ where 
\beq
\ou{0}{g}{1}{}'=h_t U^{-1} \ou{0}{h}{1}  U \quad \mbox{with} \quad U(t)=\left[\begin{matrix} \exp(\phi_0 \,\frac{i}{2}\sigma_3)  & 0 \\ 0 & \exp(\phi_0 \,\frac{i}{2}\sigma_3)\end{matrix}\right]
\eeq
Under the canonical homeomorphism to $\mc{Q}^*_0$ given above, this configuration maps to
\beq
V^{\h{i}}=(u^{-1}h u h^{-1})^{\h{i}} \quad \mbox{with}\quad u= \exp\left(\phi_0 \,\textstyle{\frac{i}{2}}\sigma_3\right) \label{Z2Generator}
\eeq
where it is understood in this expression\footnote{In this expression and throughout hatted lower cases indices like $\h{i}$ take values from 1 to 4 while unhatted lower case indices take values from 1 to 3.} that the components $V^{\h{i}}$ are extracted from the expression $V=V^4 \bm{1}+V^i \,i\sigma_i$. I claim that this is a generator of $\pi_4(\mc{G}/\mc{H})=\bb{Z}_2$. To see this, I must first digress to discuss more generally the generators of $\pi_4(\bb{S}^3)=\bb{Z}_2$ (see e.g. \cite{Finkelstein:1968hy,Giulini:1993gd,Williams:Skyrme} for a related discussion in the context of the Skyrme model).

\subsection{The Hopf map and suspensions}
The group $\pi_4(\bb{S}^3)=\bb{Z}_2$ has just two elements, so it is sufficient to find a single map taking $\bb{S}^4\rightarrow \bb{S}^3$ that is not deformable to the identity. Any map that is homotopic to this map will then also be a generator of $\bb{Z}_2$. To construct the generator, consider first the lower dimensional case of maps from $\bb{S}^3$ onto $\bb{S}^2$. Since $\pi_3(\bb{S}^2)=\bb{Z}$, this group is also generated by a single map, and the canonical generator is known as the Hopf map. This map can be thought of as a fibration of the three-sphere into a non-trivial bundle of $U(1)$ fibers over $\bb{S}^2$. The projection map of the bundle $\pi:P=\bb{S}^3\rightarrow M=\bb{S}^2$ is the Hopf map itself. Explicitly it can be constructed as follows. Consider a vector field $n^{i}$ in the tangent space of $\bb{S}^3$ whose magnitude $\delta_{ij}n^i n^j=1$ constrains it to live in an $\bb{S}^2$ submanifold of the tangent space. The group $SU(2)$ acts transitively on this space via the adjoint action $n=n^{i}\,i\sigma_i \rightarrow n' =ana^{-1}$ for any $a:\bb{S}^3 \rightarrow SU(2)$. Now, take $a$ to be the standard generator, $h$, of $\pi_3(\bb{S}^3)$ given in (\ref{Zgenerator}), and take $n^i=(0,0,1)$. Then, since $h$ is not deformable to the identity and has winding number one, the map $n'=hnh^{-1}=h \,i\sigma_3 h^{-1}$ is a map from $\bb{S}^3$ to $\bb{S}^2$ with winding number one. It therefore serves as the generator of $\pi_3(\bb{S}^2)=\bb{Z}$. This is the Hopf map.

A well known result of homotopy theory is that any {\it suspension} of the Hopf map is a generator of $\pi_4(\bb{S}^3)=\bb{Z}_2$. An example of such a suspension is any continuous map $V:\bb{S}^4\rightarrow \bb{S}^3$ such that the restriction of the map to the $\bb{S}^3$ equator of $\bb{S}^4$ reduces to the Hopf map. For example, take $\chi$ to be the azimuthal angle on $\bb{S}^4$. Then the map $V^{\h{i}}=(huh^{-1})^{\h{i}}$, where as before $u=\exp\left(2\chi \,\frac{i}{2}\sigma_3 \right)$, is such a suspension since on the equator at $\chi=\frac{\pi}{2}$, the map reduces to $V(\pi/2)=h\,i\sigma_3 h^{-1}$, the Hopf map.

\subsection{The $\bb{Z}_2$ configurations in $\mc{Q}$}
We now return to our configuration $V^{\h{i}}=(u^{-1}h u h^{-1})^{\h{i}}$. Evaluated at $t=0$, this map reduces to $-i \sigma_3 h\, i\sigma_3 h^{-1}$. This is simply the Hopf map followed by a constant rotation by $-\pi$, and is therefore homotopic to the Hopf map. In turn, the full map $V^{\h{i}}$ is homotopic to a suspension of the Hopf map in $\bb{S}^4$, and is therefore a generator of $\pi_4(\mc{G}/\mc{H})=\bb{Z}_2$. Thus, we have constructed the generator of the $\bb{Z}_2$ degeneracy in  the sector $\mc{Q}_{-1}$. Clearly a similar construction holds in the sector $\mc{Q}_{1}$. Moreover, it should be clear that any time dependent spatial rotation of the same generic form as $\varphi$ is a generator of $\bb{Z}_2$. It is slightly less obvious, but still true, that a time dependent spatial {\it translation} of de Sitter space about one full revolution is also a generator\footnote{The easiest way to see this is to picture de Sitter space as the hyperboloid embedded in $\bb{R}^{1,4}$ with coordinates $\{T,X,Y,Z,W\}$. A global $2\pi$-rotation about the $XY$ plane is clearly homotopic to a global $2\pi$-rotation about $XZ$ since any spatial plane can be continuously deformed into another. By the same token it is homotopic to a $2\pi$-rotation about the $XW$ plane. But the latter is interpreted as a translation of the spacetime about one full revolution. The same reasoning applies for time dependent rotations and translations.} of $\bb{Z}_2$.

To construct the generator in all sectors $\mc{Q}_q$ I will borrow well known results from the Skyrme model \cite{Finkelstein:1968hy,Giulini:1993gd,Williams:Skyrme}. Consider the action of $\varphi$ on the base points $\{\ou{m}{\cA}{n},\g_5\}$, where we recall $q\equiv m-n$. It can be shown that in any sector $\mc{Q}_q$ with {\it odd} charge $q$, the action of $\varphi$ on the base point $\{\ou{m}{\cA}{n},\g_5\}$ is a generator of $\bb{Z}_2$. On the other hand, the map is deformable to the trivial map in any sector with {\it even} $q$. In sectors of even charge, the generator can be constructed from exchanges of the geometries (see Fig. \ref{Exchange}).

\begin{figure}
 \begin{center}
\includegraphics[height=6.0cm]{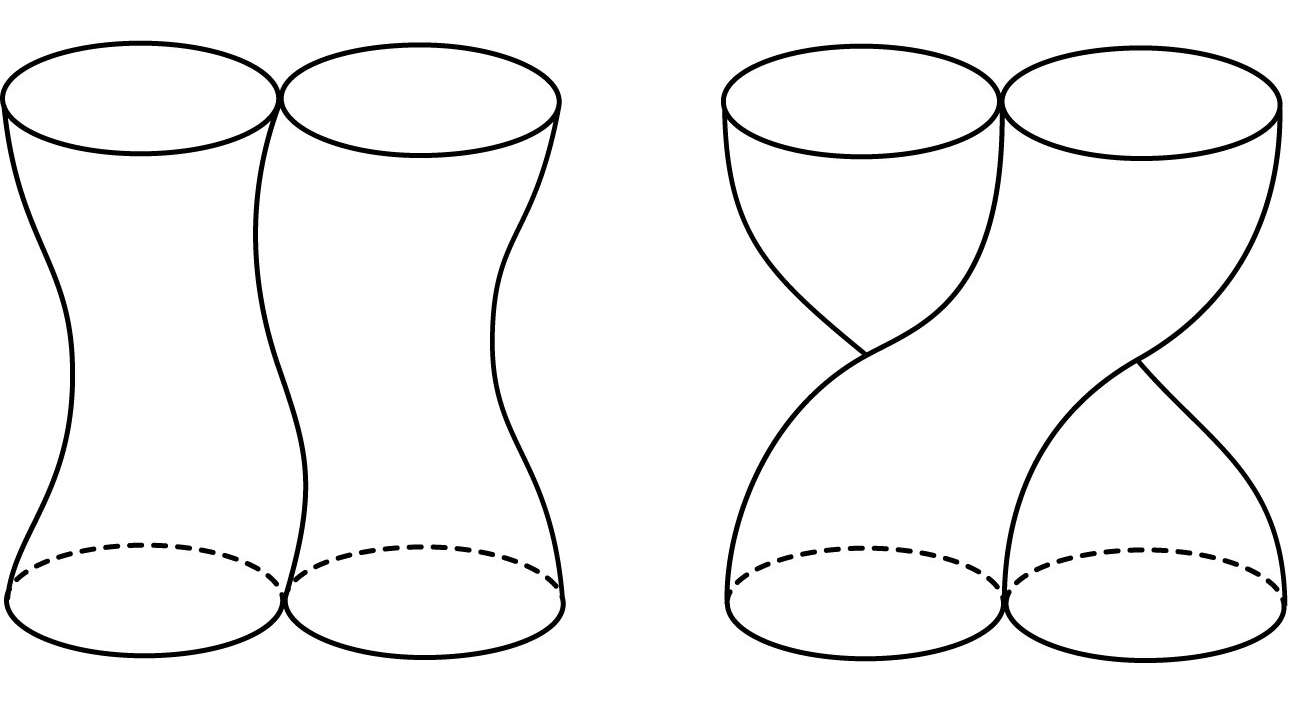}
   \end{center}
  \caption{\label{Exchange} The exchange of two copies of de Sitter space in the $q=\pm 2$ sector. This map is homotopic to $2\pi$ twist of just one of the copies, and is therefore a generator of $\bb{Z}_2$ in the $q=\pm 2$ sector.
  }
 \end{figure}

Consider for example the $q=-2$ sector whose base point consists of two copies of de Sitter space attached along a degenerate surface. A diffeomorphism representing an exchange of these two copies is homotopic to a $2\pi$ twist of just one copy. Thus, the exchange is the generator of $\bb{Z}_2$ in this case. Similar results hold in all sectors with even $q$.

\section{Fermionic Quantization in General Relativity}
I now turn to the details of the fermionic quantization procedure, focusing on the most obvious model: General Relativity. It is natural to look toward topological $BF$ theory on a $Spin(4,1)$ bundle, as the solution space contains the space of flat de Sitter connections. However, this theory does not retain enough information about the Cartan decomposition of the de Sitter connection to yield the desired results. Similarly, the ordinary Macdowell-Mansouri action does not work. However, as I will now show, ordinary Einstein-Cartan gravity restricted to the set of flat de Sitter connections does retain enough geometric structure in the covariant phase space to allow for fermionic quantization.

Consider the Einstein-Cartan action with a cosmological constant given be
\beq
S=\frac{1}{4k}\int_{M}\epsilon_{IJKL}\,e^I\,e^J\,\left(R^{KL}_\omega-\frac{1}{2\ell^2}e^K\,e^L\right)\,.
\eeq
The goal is to write this in a manifestly $Spin(4,1)$ invariant form by introducing the symmetry breaking field $V^{\hI}$. Once this is done, restriction to the set of flat connections can be achieved by introducing a Lagrange multiplier enforcing the flatness constraint. 

First consider the Cartan decomposition of the connection $\cA^{\hI\hJ}$ into the lift of the spin connection and tetrad to the $Spin(4,1)$ bundle. Recalling that the lift of tetrad is given by $e^{\hI}=\ell D_{\cA}V^{\hI}$, the decomposition is given by
\beq
\cA^{\hI\hJ}=\omega^{\hI\hJ}+2D_{\cA}V^{[\hI}V^{\hJ]}\,.
\eeq
where $\omega^{\hI\hJ}$ is the lift of the $Spin(3,1)$ connection to the $Spin(4,1)$ bundle. Its curvature is given by
\beq
R^{\hI\hJ}_{\omega}=\mc{F}^{\hI\hJ}_{\cA}+D_\cA V^{\hI}\, D_\cA V^{\hJ}+2V^{[\hI} {\mc{F}^{\hJ]}}_{\hK} V^{\hK}\,.
\eeq
The last term on the right hand side above is the lift of the torsion to the $Spin(4,1)$ bundle, and will therefore not enter into the action.

The Einstein-Cartan action can therefore be written in a manifestly $Spin(4,1)$ invariant way as
\beq
S_{EC}=\frac{\ell^2}{4k}\int_{M}\epsilon_{\hI\hJ\hK\hL\hM}V^{\hI} \, D_{\cA}V^{\hJ} \, D_{\cA}V^{\hK}\left(\mc{F}^{\hL\hM}_\cA+\frac{1}{2}D_\cA V^{\hL}\, D_\cA V^{\hM} \right)\,.
\eeq
To see that this is action is gauge equivalent to the Einstein Cartan action, one simply needs to transform to the Einstein-Cartan gauge where $V=(0,0,0,0,1)$, and identify the lift of the tetrad with $e^{\h{I}}\equiv \ell D_{\mc{A}} V^{\h{I}}$.

We eventually need to evaluate the variation of this action on the covariant phase space $\mc{Q}$. To enforce the flatness constraint, add a Lagrange multiplier term of the form
\beq
S_{BF}=\alpha \int_M B_{\hI\hJ}\,\mc{F}^{\hI\hJ}_{\cA}\,.
\eeq
Alone, this is the familiar topological $BF$ action whose solution space depends only on the cohomology class of the manifold. Since the manifold has topology $\bb{R}\times \bb{S}^3$, and one can always construct a trivial bundle over $\bb{S}^3$, we can safely fix the $Spin(4,1)$-bundle to be trivial. In this case, the first and second cohomology classes are also trivial. This means that the connection $\cA$ is always gauge related to the zero connection so that $\cA=-dg\,g^{-1}$, and that all closed two-forms are also exact.

Varying with respect to the field $B_{AB}$ enforces the constraint $\mc{F}=0$, as desired, but it still needs to verified that the term does not modify the symplectic form. I will work in the covariant phase space formulation wherein one first takes an arbitrary variation of the action evaluated on $t\in [t_i,t_f]$, and sets the bulk terms to zero by imposing the classical equations of motion. The remaining terms comprise the symplectic one-form giving rise to the conserved symplectic form.

Thus, given the action $S=S_{EC}+S_{BF}$, the variation of the action (which is taken to be arbitrary) is performed, and the result is then evaluated on the constraint surface given by $\mc{F}=0$. The result is the following:
\beqa
\dl S &=& \frac{\ell^2}{4k} \int_{\p M}\epsilon_{\hI\hJ\hK\hL\hM}\,V^{\hI} \,D_\cA V^{\hJ}\, D_\cA V^{\hK}\,\dl \cA^{\hL\hM} \\
& & -\frac{\ell^2}{2k} \int_{\p M} \epsilon_{\hI\hJ\hK\hL\hM}\, V^{\hI} \,D_\cA V^{\hJ}\,D_\cA V^{\hK} \, D_\cA V^{\hL} \,\dl V^{\hM} \\
& & \alpha \int_{\p M} B_{\hI\hJ} \, \dl \cA^{\hI\hJ} \\
& & -\alpha \int_{M} D_\cA B_{\hI\hJ} \,\dl \cA^{\hI\hJ}\,.
\eeqa
Since I have made the assumption that the bundle is trivial over $\bb{R}\times \bb{S}^3$ and the connection is flat, the equation of motion $D_{\cA} B_{\hI\hJ}=0$ implies $B_{\hI\hJ}=D_{\cA}\lambda_{\hI\hJ}$ for some Lie algebra valued one-form $\lambda_{\hI\hJ}$. Inserting this into the expression on the third line above, we have
\beq
\alpha \int_{\p M} B_{\hI\hJ}\, \dl \cA^{\hI\hJ}=\alpha \int_{\p M}\lambda_{\hI\hJ}\, D_{\cA}\dl \cA^{\hI\hJ}\,.
\eeq
The symplectic one-form on the covariant phase space is then found be restricting the variations in the boundary integrals to zero. Since $D_{\cA}\dl \cA^{\hI\hJ}=\dl \mc{F}^{\hI\hJ}=0$ on the solution space, the expression above is identically zero. Thus, the symplectic one form on the covariant phase space reduces to
\beqa
\bm{J} &=& \frac{\ell^2}{4k} \int_{\bb{S}^3}\epsilon_{\hI\hJ\hK\hL\hM}\,V^{\hI} \,D_\cA V^{\hJ}\, D_\cA V^{\hK}\,\dl \cA^{\hL\hM} \\
& & -\frac{\ell^2}{2k} \int_{\bb{S}^3} \epsilon_{\hI\hJ\hK\hL\hM} \,V^{\hI} \,D_\cA V^{\hJ}\,D_\cA V^{\hK} \, D_\cA V^{\hL} \,\dl V^{\hM} \,.
\eeqa
The conserved symplectic form on the covariant phase space is therefore
\beqa
\bm{\Omega} = -\dl \bm{J} &=&- \frac{\ell^2}{4k} \int_{\bb{S}^3}\dl (\epsilon_{\hI\hJ\hK\hL\hM}\,V^{\hI} \,D_\cA V^{\hJ}\, D_\cA V^{\hK})\ \dl \cA^{\hL\hM} \\
& & +\frac{\ell^2}{2k} \int_{\bb{S}^3} \dl (\epsilon_{\hI\hJ\hK\hL\hM} \,V^{\hI} \,D_\cA V^{\hJ}\,D_\cA V^{\hK} \, D_\cA V^{\hL}) \ \dl V^{\hM} \,.
\eeqa
In this expression, it is understood that $\dl$ is the exterior derivative on the covariant phase space, and wedge products between differential forms are assumed. Here and throughout, differential forms on the phase space will be written in $\textbf{bold}$ font. In fact this expression can be simplied even further by performing a gauge fixing. Recall that the connection $\cA$ is gauge related to the flat connection. As the above expression for the symplectic form is gauge invariant, one can always fix the gauge to the trivial gauge where $\cA^{\hI\hJ}=0$ and require that all variations are such that $\dl \cA^{\hI\hJ}=0$. Up to a constant element of $Spin(4,1)$, this completely fixes the gauge. The symplectic form then becomes
\beqa
\bm{\Omega} &=& \frac{\ell^2}{2k} \int_{\bb{S}^3} \dl (\epsilon_{\hI\hJ\hK\hL\hM} \,V^{\hI} \,d V^{\hJ}\,d V^{\hK} \, d V^{\hL})\ \dl V^{\hM} \\
&=& -\frac{5\ell^2}{4k} \int_{\bb{S}^3} \epsilon_{\hI\hJ\hK\hL\hM} \,\dl V^{\hI} \,\dl V^{\hJ}\, d V^{\hK} \, d V^{\hL} \, d V^{\hM}
\eeqa
We now come to an important result. Since both $\dl V^{\hI}$ and $dV^{\hJ}$ are perpendicular to $V^{\hI}$ itself, but the contraction of the alternating symbol is zero unless it is contracted onto a set of five linearly independent vector fields, the symplectic form is identically zero
\beqa
\bm{\Omega}=0. 
\eeqa
In retrospect this result should have been expected. It is a reflection of the topological nature of the theory, and indicates that the topological degrees of freedom have been isolated. When the symplectic form is pulled back to the constraint surface formed by the flatness constraint, there are no local degrees of freedom. Thus, all vectors in the tangent space of the constraint surface are the generators of gauge transformation and therefore Lie drag the symplectic form so that $\mathcal{L}_{\bm{\bar{V}}} \bm{\Omega}=\dl \bm{\Omega(\bar{V})}=0$. Since this is true for every vector in the tangent space of the constraint surface, the symplectic form must be zero.

Despite the symplectic form vanishing, one can identify a phase space from the symplectic one-form, $\bm{J}$, which is unequivocally non-zero. As I will show, the symplectic one-form contains important topological information about the phase space. Performing a maximal gauge-fixing to the Cartan gauge, we have
\beq
\bm{J}=-\frac{\ell^2}{2k} \int_{\bb{S}^3} \epsilon_{\hI\hJ\hK\hL\hM} \,V^{\hI} \,d V^{\hJ}\,d V^{\hK} \, d V^{\hL}\, \dl V^{\hM} \,.
\eeq
It is clear from this form that both the position and momentum variables are completely fixed by the field $V^{\hI}$. I conclude that the {\it phase space}, $\mc{P}$, corresponding to Einstein-Cartan gravity in the $Spin(4,1)$ invariant Cartan form is given by the set of smooth maps $V:\bb{S}^3\rightarrow \mc{G}/\mc{H}\simeq \bb{R}\times \bb{S}^3$. Thus, the phase space has the key property $\pi_1(\mc{P})=\pi_4(\mc{G}/\mc{H})=\bb{Z}_2$. Although this is a property of the phase space and it is not known if it holds through a polarization to a configuration space, I will now show that, because of the topological nature of the theory, the property holding on the phase space alone  is sufficient to allow for fermionic quantization.

\subsection{Geometric quantization}
Geometric quantization provides the most convenient route to understanding the fermionic properties of our model. In this framework, one introduces a complex line-bundle over the phase space which plays the role of a pre-quantum Hilbert space. One then defines a $U(1)$ connection whose associated curvature is proportional to the symplectic form. To obtain the true Hilbert space, one must implement a polarization on the pre-quantum Hilbert space that restricts attention to wavefunctions on a choice of configuration space. In our case, since the theory is topological, the subtleties of defining a polarization are inconsequential -- the full wavefunctional must be covariantly constant in any direction in the phase space. Instead we must pay close attention to the topological properties of the phase space.

First note that he connection is always only unique up to the addition of a closed and exact one-form since $\bm{J}$ and $\bm{J}+\dl f$ define the same symplectic form. As this is simply a $U(1)$ transformation, this amounts to working in a different gauge thereby defining a canonical transformation. However, in the case that the cohomology class of the phase space is non-trivial, one can have closed one-forms that are not exact. As I have shown, $\pi_1(\mc{P})=\pi_4(\mc{G}/\mc{H})=\bb{Z}_2$. I can make a slightly stronger statement by identifying the path connected components of the phase space, $\mc{P}$. The phase space splits into disjoint $\bb{Z}$-sectors, $\mc{P}_q$, characterized by the winding number of the map $V:\bb{S}^3\rightarrow \mc{G}/\mc{H}\simeq \bb{R}\times \bb{S}^3$. Each of these sectors is path connected and they are all homeomorphic to each other. Thus we have $\pi_1(\mc{P}_q)=\bb{Z}_2$. For a path connected space with an abelian fundamental group, the first homology group can be identified with the fundamental group itself. 
Thus, we have $\mc{H}_1(\mc{P}_q)=\pi_1(\mc{P}_q)=\bb{Z}_2$. 

The cohomology group, $\mc{H}^1(\mc{P})$, is the dual of the homology group, and from de Rham's Theorem (see \cite{Nakahara}) must be non-trivial. In fact, as we show in the Appendix \ref{Cohomology}, we have $\mc{H}^1(\mc{P})=\bb{R}$. This means that there are closed, non-exact one-forms over $\mc{P}_q$.

The symplectic one-form $\bm{J}$ is in fact exact (see Appendix {\ref{Appendix}} for the proof of this). So let $\bm{\mc{K}}$ be a closed non-exact one-form (i.e. $\dl \bm{\mc{K}}=0$ but $\bm{\mc{K}}\neq \dl{\lambda}$). Consider the line integral $\int_\gamma \bm{\mc{K}}$ around any closed loop $\gamma$ serving as a generator of $\pi_1(\mc{P})=\bb{Z}_2$. Since the one-form is closed, the integral is invariant under continuous deformations of the curve $\gamma$. More specifically, the integration can be viewed as a non-degenerate map $\int:\mc{H}_1(\mc{P})\times \mc{H}^1(\mc{P})\simeq \bb{Z}_2\times \bb{R} \rightarrow \bb{R}$. Putting all this together, we have the generic result
\beq
\int_\gamma \bm{\mc{K}}=z\theta
\eeq
where $\theta\in\bb{R}$ and $z=0$ if $\gamma$ is contractable to the identity, and $z=1$ if $\gamma$ is a generator of $\pi_1(\mc{P})=\bb{Z}_2$. One can always normalize the one-form $\bm{\mc{K}}$ so that $\theta=\pi$. Assuming that $\bm{\mc{K}}$ is normalized as such, we have
\beq
\exp\left(i \int_\g \bm{\mc{K}}\right)=\begin{cases} \ \  1 &  \mbox{if $\g$ is contractable to the trivial map} \\
-1 & \mbox{if $\g$ is a generator of $\pi_1(\mc{P})=\bb{Z}_2$.} \end{cases}
\eeq

Now consider a wavefunction on the {\it pre-quantum} Hilbert space $\Phi[V]$. Take the connection to be $\bm{J'}=\bm{J}+\bm{\mc{K}}$ so $\bm{\mc{D}_{J'}}=\dl-i\bm{J}-i\bm{\mc{K}}$
\beq
\Psi[V]=\exp \left( i \int^V_{V_0,\gamma} \bm{J'} \right) \Phi[V]
\eeq
where $V_0$ is some reference configuration\footnote{Since the integral around a close loop is invariant under homotopy preserving deformations of the loop, the precise configuration of the reference point $V_0$ is inconsequential.} and the path, $\g$, in the integral is taken to be any smooth path in the phase space connecting $V_0$ and $V$ . Since $\dl \bm{\mc{K}}=0$, the integral is invariant under small deformations that preserve the end points of the path. From the identity\footnote{It may seem peculiar that $\dl \exp\left(i\int^V_{V_0}\bm{\mc{K}}\right)=i\bm{\mc{K}}$ as this would seem to imply that $\bm{\mc{K}}$ is exact. However, in a local, contractable neighborhood of the base point $V_0$, all closed forms are also exact, and this is simply a reflection of this fact.}
\beq
\bm{\mc{D}_{J'}}\Psi=\exp \left( i \int^V_{V_0, \g} \bm{J'} \right) \dl \Phi \label{DD}
\eeq
strong restrictions can be imposed on the wave functional. Since the symplectic form is identically zero, every vector in the phase space is tangent to a gauge orbit, along which the wavefunctional must be {\it covariantly} constant. Thus, given any vector field $\bm{\bar{W}}$ in the phase space, we must have
\beq
\bm{\iota_{\bar{W}}}\bm{\mc{D}_{J'}}\Psi =0. \label{Zero}
\eeq 
However, from (\ref{DD}), this implies $\bm{\iota_{\bar{W}}}\dl\Phi =0$. Since this is true of any vector, we must have $\dl \Phi=0$. Therefore $\Phi$ cannot depend on local information on the phase space, but only on global or topological properties. Since the phase space is given by the set of maps $V:\bb{S}^3\rightarrow \mc{G}/\mc{H}$, it is completely characterized by the homotopy group $\pi_3(\mc{G}/\mc{H})=\bb{Z}$. Thus, the functional $\Phi$ can only depend on the winding number $q$ of the map in each sector $\mc{P}_q$. This is precisely analogous to the topological $n$-sectors of QCD. Following in stride, one can build a Hilbert space of topological $q$-states denoted $\phi_q$, with inner product $\langle \phi_{q'}|\phi_q \rangle =\delta_{q'q}$. Restricting to the $q$ sector, up to an overall phase, we have
\beq
\Psi_q=\exp \left(-i \int^V_{V_0, \g} \bm{\mc{K}} \right) \phi_q \,.
\eeq

We now come to the conclusion of this argument. Consider the evolution of the wave function $\Psi[V]=\exp \left(-i \int^V_{V_0, \g} \bm{\mc{K}} \right) \Phi_{q}$ along a generator of $\pi_1(\mc{P}_q)=\bb{Z}_2$. Any generator will work, but for definiteness consider the evolution of the wavefunctional in the de Sitter sector ($q=-1$) under the $2\pi$-rotational diffeomorphism given above. $\Phi[V]=\Phi_{-1}$ is single valued along the curve (in fact it is constant), and the pre-factor $\exp \left(-i \int^V_{V_0, \g} \bm{\mc{K}} \right)$ is a pure phase. Thus, at any point along $\gamma$, we have $\Psi[V]=e^{i\theta}\phi_{-1}$. At the end point, the curve is a closed, non-contractable loop that is a $\bb{Z}_2$ generator. Thus, after a $2\pi$ rotation, the pre-factor is $-1$. In total, under a transformation representing a physical rotation by $2\pi$, the wavefunctional $\Psi[V]$ transforms according to 
\beqa
\Psi[V] \ \Longrightarrow \ -\Psi[V]
\eeqa


\subsection{Polarization} Although all of our discussion has been about wavefunctionals on the prequantum Hilbert space, due to the topological nature of the theory, the states $\Psi_q$ are already naturally polarized. To see this, recall that a polarization is implemented on the prequantum Hilbert space by restricting to wavefunctionals that are {\it covariantly} constant along a chosen direction. A polarization of the pre-quantum Hilbert space is given by a set of $n$ linearly independent vector fields $\bm{\bar{Z}^{(i)}}$ on the (locally) 2n-dimensional phase space $\mc{P}$ such that $\bm{\Omega(\bar{Z}^{(i)},\bar{Z}^{(j)})}=0$. The polarization is then implemented in a $U(1)$ covariant manner by
\beq
\bm{\iota_{\bar{Z}^{(i)}}\mc{D}_{J'}}\Psi=0.
\eeq
From (\ref{Zero}), this condition is automatically satisfied. This is simply a reflection of the topological nature of the theory.

\subsection{Summary}
To summarize, we constructed a topological quantum theory that is at once rich enough to embed de Sitter space in a full quantum Hilbert space, yet simple enough to elucidate the topological nature of the theory. The phase space of Einstein Cartan theory in the Cartan formulation of the set of flat de Sitter connection on $\bb{R}\times \bb{S}^3$ is given by the set of maps $V:\bb{S}^3\rightarrow \bb{R}\times \bb{S}^3$. This phase space has a non-trivial fundamental group given by $\pi_1(\mc{P})=\pi_4(\bb{R}\times \bb{S}^3)=\bb{Z}_2$. Thus, there are non-contractable loops in the phase space that cannot be deformed to the identity. I have shown explicitly that in the de Sitter sector ($q=-1$), one such non-contractable loop is a parameretized $2\pi$ rotation. 

In addition, it has been demonstrated that the cohomology group of the phase space is non-trivial. This gives rise to a quantization ambiguity. In addition to the standard quantization of the theory (the $\Phi$-representation) where diffeomorphisms act trivially in the quantum theory, there exists an alternative quantization (the $\Psi$ representation) where diffeomorphisms act {\it projectively}. In this representation, under a transformation representing a generator of $\pi_1(\mc{P})$, the wavefunctional transforms according to $\Psi \longrightarrow -\Psi$. Thus, the de Sitter geometry itself can be quantized fermionically.

\section{Concluding Remarks}
In ordinary quantum field theory, it is generally taken as given that bosons are comprised of ordinary commuting fields, and fermions are comprised of Grassman fields. However, the non-linear structure of certain field theories admits an alternative means for fermionic statistics to emerge. Here I have given strong evidence that this possibility may be realized in quantum gravity. Together with our previous work on emergent fermions in torsional systems \cite{Randono:2010cd,Randono:2011bb}, we now have two examples of fermionic geometries. Regarding the present work, the prospect that an entire spacetime could itself behave fermionically challenges many implicit tenets of quantum gravity, quantum cosmology, and even quantum field theory. Furthermore, the spacetime geometry in question is hardly exotic, being the simplest model spacetime beyond Minkowski space. Thus, the physical ramifications of fermionic geometries need to be explored in depth.

\appendix

\section{Clifford algebra conventions}
Here I will review the basic conventions used in this paper regarding the Clifford algebra notation. The Clifford algebra is spanned by a set of Clifford matrices $\{\g^I\}$ satisfying $\g^I \g^J +\g^J\g^I=2\eta^{IJ}$ where $\eta^{IJ}=diag(-1,1,1,1)$. The volume element $\g_5$ is defined to be $\g_5=\frac{i}{4!}\epsilon_{IJKL}\g^I\g^J\g^K\g^L=i\g^0\g^1\g^2\g^3$, where it is understood that the alternating symbol $\epsilon_{IJKL}$ is such that $\epsilon_{0123}=-\epsilon^{0123}=1$.
The ten dimensional de Sitter algebra, $\mf{spin}(4,1)$ is spanned by the basis elements $\{\frac{1}{2} \g^{[I}\g^{J]}\,,\, \frac{1}{2}\g_5 \g^K \}$. The Lorentz stabilizer subalgebra, denoted $\mf{h}=\mf{spin}(3,1)$ is spanned by $\{\frac{1}{2}\g^{[I}\g^{J]}\}$ and its complement $\mf{p}$ in $\mf{spin}(4,1)$ is spanned by $\{\frac{1}{2}\g_5 \g^K\}$. Given an antisymmetric Lorentz valued matrix $A^{IJ}=A^{[IJ]}$, the index free object $A$ is given by $A\equiv \frac{1}{4}\g_{[I}\g_{J]}\,A^{IJ}$.

The $\mf{spin}(4,1)$ connection coefficient in a local trivialization denoted by $\mc{A}$ splits correspondingly. However, it is convenient to represent the tetrad as simply $e\equiv \frac{1}{2} \g_I \, e^I$ which pulls the $\g_5$ out of the decomposition $\cA=\omega \oplus \frac{1}{\ell} \g_5 \,e$.

\section{Generators of $\mc{H}_1(\mc{P})$\label{Appendix}} 
We have shown that the phase space of Einstein-Cartan gravity in the gauge gravity framework has the property $\mc{H}_1(\mc{P})=\bb{Z}_2$, indicating that there are closed, but non-exact one-forms in the cotangent bundle. Here we show that the symplectic one-form $\bm{J}$ is {\it not} a generator of $\bb{Z}_2$. We first note that given a closed, non-contractable loop $\gamma$ in $\mc{P}$ (i.e. a generator of $\pi_1(\mc{P})$), If $\bm{\mc{K}}$ is a generator of $\mc{H}_1(\mc{P})$, then 
\beq
\int_\gamma \bm{\mc{K}} \neq 0\,.
\eeq
We will now show that the line integral of $\bm{J}$ is always identically zero. 

Since $\mc{P}$ splits into topologically distinct sectors $\mc{P}_q$, all of which are homeomorphic to each other, it is sufficient to consider a generator, $\gamma$, of $\pi_1(\mc{P}_0)=\bb{Z}_2$. Via the homeomorphism, this curve can be mapped into all other sectors and its topological properties will be preserved. Thus, let $\gamma:t\in[t_i,t_f]\rightarrow \mc{P}_0$ be such a generator. Since all such generators are homotopic to each other, we can take $\gamma$ to be the map (\ref{Z2Generator}). Now consider the line integral of $\bm{J}$
\beq
\int_\g \bm{J} =-\frac{\ell^2}{2k} \int^{t_f}_{t_i} dt \int_{\bb{S}^3} \epsilon_{\hI\hJ\hK\hL\hM} \,V^{\hI} \,d V^{\hJ}\,d V^{\hK} \, d V^{\hL}\, \frac{\p V^{\hM}}{\p t}\,.
\eeq
Since $V^{hI}(t_i)=V^{hI}(t_i)=(0,0,0,0,1)$ is constant on $\bb{S}^3$, we can formally compactify the integral about the loop as an integral over $\bb{S}^4$:
\beq
\int_\g \bm{J} \sim \int_{\bb{S}^4} \epsilon_{\hI\hJ\hK\hL\hM} \,V^{\hI} \,d V^{\hJ}\,d V^{\hK} \, d V^{\hL} \, d V^{\hM}\,.
\eeq
This is to be evaluated on the configuration (\ref{Z2Generator}). However, for this configuration $V^{\h{0}}$ and $d V^{\h{0}}$ is everywhere zero, so there are simply not enough non-zero components to saturate the alternating tensor in the integrand. Therefore the integral is identically zero, so $\bm{J}$ is {\it not} a generator of $\mc{H}_1(\mc{P})$.
 
\section{Proof that $\mc{H}^1(\mc{P})=\bb{R}$\label{Cohomology}}
The de Rham theorem states that the cohomology group $\mc{H}^n(M)$ is the vector space dual of the homology group $\mc{H}_n(M)$ and that the map $\Lambda:\mc{H}_n(M)\times\mc{H}^n(M)\rightarrow \bb{R}$ given by
\beq
\Lambda(\g,\omega)=\int_\g \omega 
\eeq
where $\g\in \mc{H}_n(M)$ and $\omega \in \mc{H}^n(M)$ is {\it bilinear} and {\it non-degenerate}. Thus, since $\mc{H}_1(\mc{P})=\pi_1(\mc{P})=\bb{Z}_2$, it must be that $\mc{H}^1(\mc{P})$ is non-trivial as well. This means that there exists closed, non-exact one-forms, $\bm{\omega}$ whose line integral $\int_\g \bm{\omega}\neq 0$ around a closed but non-contractable curve $\g$. Let $\bm{\omega}_1$ and $\bm{\omega}_2$ be two such one-forms. Since $\omega$ is closed, the integral is homotopy invariant meaning $\int_{\g'}\bm{\omega}=\int_{\g}\bm{\omega}$ if $\g'$ is homotopic to $\g$ (i.e. it is a generator of $\bb{Z}_2$). 

Suppose we have two elements of $\mc{H}^1(\mc{P})$, $\bm{\omega'}$ and $\bm{\omega}$ and consider the difference $\bm{\omega'}-a\bm{\omega}$ where
\beq
a=\frac{\int_\g \bm{\omega'}}{\int_\g \bm{\omega}}\,.
\eeq
As the set of one forms is a vector space over the reals, generically $a\in \bb{R}$. Now, given any non-contractable loop, $\g'$, we have
\beqa
\int_{\g'}\bm{\omega'}-a\bm{\omega}&=&\int_{\g'}\bm{\omega'} -\frac{\int_\g \bm{\omega'}}{\int_\g \bm{\omega}}\int_{\g'}\bm{\omega} \\
&=& \int_{\g'}\bm{\omega'} -\frac{\int_{\g'} \bm{\omega'}}{\int_{\g'} \bm{\omega}}\int_{\g'}\bm{\omega} \\
&=&0\,.
\eeqa
Thus, $\bm{\omega'}$ and $a\bm{\omega}$ must be equivalent up to an exact form
\beq
\bm{\omega'}=a\bm{\omega}+\dl f\,.
\eeq
Since $a$ is a generic real number we must have $\mc{H}^1(\mc{P})=\bb{R}$.

\cite{Friedman:1982du}

\section*{Acknowledgment}
This work was in part supported in part by the NSF International Research Fellowship grant OISE0853116.

\bibliography{FermionicdS3}

\end{document}